\documentclass[article]{aa}
\pdfoutput=1
\usepackage{amsmath}
\usepackage{txfonts}
\usepackage{graphicx}
\usepackage{xcolor}
\usepackage[utf8]{inputenc}
\usepackage[parfill]{parskip}

\begin{document}

\renewcommand{\labelitemi}{$\circ$}
\newcommand{\sud}[1]{{\color{red} #1}}
\newcommand{\Fried}{\textbf{$r_{0}$}}
\newcommand{\rephrase}[2]{}
   \title{Correction of atmospheric stray light in restored slit spectra}
   \titlerunning{Stray light correction in slit spectra}

   \author{S. Saranathan\inst{1,2},
           M. van Noort\inst{1},
           S. K. Solanki\inst{1,3}
          }

   \institute{Max-Planck Institut f{\"u}r Sonnensystemforschung,
              Justus-von-Liebig-Weg 3, 37077 G{\"ottingen}, Germany
         \and
              Georg-August-Universit\"at, Friedrich-Hund-Platz 1, 37077 G\"ottingen, Germany
         \and
              School of Space Research, Kyung Hee University, Yongin, Gyenoggi-Do 446-701, Republic of Korea
             }

   \date{}


  \abstract
   {A long-standing issue in solar ground-based observations has been the contamination of data due to stray light, which is particularly relevant in inversions of spectropolarimetric data.}
   {We aim to build on a statistical method of correcting stray-light contamination due to residual high-order aberrations and apply it to ground-based slit spectra.}
   {The observations were obtained at the Swedish Solar Telescope, and restored using the multi-frame blind deconvolution restoration procedure. Using the statistical properties of seeing, we created artificially degraded synthetic images generated from magneto-hydrodynamic simulations. We then compared the synthetic data with the observations to derive estimates of the amount of the residual stray light in the observations. In the final step, the slit spectra were deconvolved with a stray-light point spread function to remove the residual stray light from the observations.}
   {The RMS granulation contrasts of the deconvolved spectra were found to increase to approximately 12.5\%, from 9\%. Spectral lines, on average, were found to become deeper in the granules and shallower in the inter-granular lanes, indicating systematic changes to gradients in temperature. The deconvolution was also found to increase the redshifts and blueshifts of spectral lines, suggesting that the velocities of granulation in the solar photosphere are higher than had  previously been observed.}
   {}

   \maketitle{}

\section{Introduction}
    \label{Sec.1}

    The degradation of observations due to stray light has been a long-standing issue in solar physics. The term stray light, in recent literature, has come to refer to extraneous undesirable contributions to pixel intensities in images and spectra caused by one or more independent physical processes. Stray light is currently understood to be the sum of multiple components with possibly differentiable characteristics (see \citet{Pillet92, Beck11} for a comprehensive review). For ground-based observations, these include the refraction of light due to seeing, as well as large-scale scattering due to the presence of aerosols, dust, molecules, etc. The former effect varies rapidly with time and is the dominant effect over distances typically up to a few arcseconds, while the latter exhibits slower modulation and brings contributions from sources that are much farther away. For example, the inexplicable brightness of sunspot umbrae \citep{Zwaan65} was found to be due to atmospheric scattering of light from the full solar disk. Stray light is also generated within telescopes and instruments by the diffraction of light, as well as scattering due to the surface roughness of optical elements. It is even present in space-based observations and may be anisotropic and inhomogeneous over the field of view (FOV). \par

    Stray-light contamination manifests itself differently in images and spectra. Images suffer a loss of spatial resolution, as well as a reduction in intensity contrasts - an effect verified by observations obtained through both space-based and ground-based instruments \citep[see][]{Uitenbroek07, Sanchez00, Danilovic08}. While this contrast reduction has been satisfactorily modelled in space-based observations \citep[e.g.][]{Mathew09, Wedemeyer09}, a discrepancy still exists in ground-based observations \citep{Scharmer10}. The effects of stray light on spectra are revealed by inversions of spectropolarimetric data. The mixing of intensities alters the depth, broadening, and symmetry of spectral lines, introducing different errors in different spectral lines. The quality of the fit in inversions is therefore generally poorer, and the reliability of the retrieved physical parameters is reduced. Several prescriptions to tackle spectral contamination can be found in the literature. \par

    \citet{OrozcoMay07} first proposed such a method by constructing a local stray-light profile from a 1\arcsec region surrounding every pixel. The inversion code was then allowed to fit the weight of this profile in the pixel being inverted. Although a significant step forwards, the method is formally incorrect as the stray-light contribution in a pixel cannot be directly estimated from the contaminated data. An improvement over this method through a self-consistent coupled inversion scheme was proposed by \citet{vanNoort12}. However, the method requires prior knowledge of the transfer functions that do not vary within the FOV. These constraints are likely not satisfied by ground-based observations (especially from spectrographs), and the preferred approach, therefore, is to restore the data as perfectly as possible before inversions.\par

    Restoration of degraded observations is only possible with an estimate of the transfer function of the optical system. Systematic efforts have been undertaken along these lines. The earliest such attempts were made by \citet{Wanders34}, \citet{Stumpff61}, and \citet{David62}. They defined an analytical point spread function (PSF) that quantified the extraneous contribution to intensity in a given pixel, as a function of distance. Subsequent theoretical and empirical refinements to the model were made by \citet{Zwaan65}, \citet{Staveland70}, \citet{Mattig71}, and later \citet{Pillet92}. Although instructive, early models of the spread function were limited in their ability to improve the quality of ground-based observations due to the use of coarse assumptions. A significant boost came from parallel developments in data acquisition, adaptive optics (AO), and image restoration techniques, such as speckle \citep{Keller92}, phase-diversity \citep{Paxman92, Lofdahl94}, and multi-frame blind deconvolution (MFBD; van Noort et al.2005), which attempt to model the true PSF as it changes with time. In the current paradigm, AO systems and image restoration methods are routinely employed to circumvent degradation and/or reverse it to an acceptable level. \par

    Despite the progress, persistent signs of degradation in ground-based observations have prompted a careful search for the sources of residual stray light. \citet{Scharmer10} found that high-order wave-front aberrations - corresponding to the wings of the PSF, which were systematically underestimated by MFBD - could be partly responsible for residual stray light in ground-based observations. They used a statistical method based on Kolmogorov statistics and knowledge of the Fried parameter \Fried{}~\citep{Fried66} to approximate corrections to the transfer functions returned by MFBD and subsequently reduce the amount of residual stray light in observations. We propose a refinement to this method by using the values of the Fried parameter to generate fully turbulence-degraded synthetic observations. Restoring the synthetic observations along with the actual observations in parallel, using MFBD, allows us to self-consistently quantify the efficiency with which the AO reversed seeing-induced degradation in the actual observations. Given that AO systems are not perfect, these efficiencies can be used to arrive at estimates of the residual degradation present in the AO-corrected, MFBD-restored observations. In this paper, we discuss the theory of the method and its implementation on spectropolarimetric data obtained from the Swedish Solar Telescope \citep[SST,~][]{Scharmer03}.

\section{Residual degradation in ground-based observations}
    \label{Sec.2}

    When observing astrophysical objects through a telescope of finite aperture, diffraction sets a fundamental limit to the spatial resolution of the final image(s). In the presence of atmospheric seeing, or due to the instruments themselves, additional degradation may be introduced. For a given optical system, quantification of image degradation in the observations is made with its transfer function. If the optical system includes the atmosphere, such as is the case for ground-based observations, the transfer functions are also time-dependent. We write the image formation model in the fourier domain:

    \begin{equation}
       \hat{I}(\vec{k}, t) = \hat{T}_{atm}(\vec{k}, t)\hat{T}_{inst}(\vec{k})\hat{O}(\vec{k}) + \hat{N}(\vec{k}),   \label{1}
    \end{equation}

    \noindent where~$\vec{k}$ and~$t$ denote spatial frequency and time, respectively, $\hat{T}_{atm}(\vec{k}, t)$ is the transfer function of the atmosphere, $\hat{T}_{inst}(\vec{k})$ is the transfer function of the telescope and relevant instruments, and $\hat{O}(\vec{k})$ represents the true solar scene. $\hat{N}(\vec{k})$ represents additive noise, and $\hat{I}(\vec{k},t)$ is the flat-fielded, dark-current subtracted image, recorded by the detector. In ground-based observations, the explicit dependence of $\hat{O}(\vec{k})$ on time is dropped because the timescales of solar evolution are much larger than the fluctuation timescales of the Earth's atmosphere. \par

    To model $\hat{T}_{atm}(\vec{k}, t)$, the approximation that it is the result of a pure phase error in the wave front is commonly made. Wave-front phase errors are introduced either by the optical elements themselves (static) or by the atmosphere (dynamic). The $\hat{T}_{atm}(\vec{k}, t)$ is then computed using:

    \begin{align}
      \zeta(\vec{x},t) &= \exp{(i\phi(\vec{x},t))},     \label{2}\\
      \hat{T}_{atm}(\vec{k},t) &= R(\zeta, \zeta^{*}),    \label{3}
    \end{align}

    \noindent where $\phi(\vec{x},t)$, $\zeta(\vec{x},t)$, and $R(\zeta, \zeta^{*})$ are the wave fronts, the pupil function, and the auto-correlation function, respectively. A perfectly flat wave front corresponds to an ideal transfer function that drops nearly linearly with $\vec{k}$. In the presence of aberrations, the decline with $\vec{k}$ is much steeper, forming images with a reduced signal to noise ratio (S/N) at high spatial frequencies. To prevent this, an AO system can be deployed combining a wave-front sensor (WFS) with a deformable mirror (DM) to measure and compensate wave-front aberrations in real time. The efficacy of such systems is a function of the number of actuators driving the DM, sampling of the wave front by the WFS, and the bandwidth of the AO. A review on the design and performance of AO systems is found in \citet{Rimmele11}. \par

    Aberrations are often quantified by projecting the corresponding spatial distribution of the wave-front phase error onto an ordered set of basis modes (Zernike polynomials, Karhunen-Lo\`eve functions, etc.):

    \begin{equation}
      \phi(\vec{x},t) = \sum_{l=0}^{\infty}{C_{l}~Z_{l}(\vec{x},t)}, \label{4}
    \end{equation}

    \noindent where $l$ denotes the mode number, $\phi$ is the wave-front phase error, $Z_{l}$ are the basis modes, and $C_{l}$ are mode amplitudes. In closed-loop operation, AO systems act as high-pass filters \citep{Rimmele11}. Specifically, the amplitudes of low-order aberrations are suppressed better than the high orders. The meaning of "low" and "high" in this context is somewhat arbitrary as no such sharp boundary exists. Nonetheless, it is instructive to rely on such a classification for the purposes of motivating what follows. The action of AO on the wave-front error can then be written as:

    \begin{equation}
      \phi^{AO} = \sum_{l=0}^{\infty}(1 - \epsilon^{AO}_{l})~C_{l}~Z_{l},  \label{5}
    \end{equation}

    \noindent where explicit dependence on $\vec{x}$ and $t$ is dropped for brevity. $\phi^{AO}$ is the phase-error after compensation by the DM, and $\epsilon^{AO}_{l}$ quantifies the suppression of mode $l$. That is, $\epsilon^{AO}_{l}$ characterises the efficiency of the AO in compensating for phase errors with a spatial distribution according to mode $l$. For perfect compensation, $\epsilon^{AO}_{l} = 1$, and the residual wave-front error would be identically zero over the entire aperture. Realistically, $\epsilon^{AO}_{l}$ would depend on $l$ and typically vary between zero and one. Given that AO systems act as high-pass filters, $\epsilon^{AO}_{l}$ would be close to one at the lower orders and close to zero at the higher orders. One would also expect the AO to become completely ineffective beyond a certain point, that is:
    
    \begin{equation}
        \epsilon^{AO}_{l} = 0, \hspace{10pt}\forall \hspace{10pt}l > L.  \label{6}
    \end{equation}
    
    \noindent However, during the application of our method to actual observations, several scenarios were observed where $\epsilon^{AO}_{l}$ was non-zero, even when $l > 200$. Further interpretations of this characteristic are provided in Sec.~(\ref{Sec.7}) of this paper. Additional complications concerning the nature of $\epsilon^{AO}_{l}$ are that they may potentially vary with seeing, and within the FOV. The dependence on seeing is due, in part, to the way in which the WFS detects the shape of incident wave fronts. For instance, a Shack-Hartmann WFS, such as the one used at SST, simultaneously images a local region in the FOV through an array of sub-apertures. Cross-correlations of the sub-images are then used to compute their relative shifts, and subsequently the local tip-tilt of wave fronts over the sub-apertures. The reliability of these shifts is particularly impacted when the seeing is very poor. Correspondingly, the $\epsilon^{AO}_{l}$ would be low. The second complication - dependence of $\epsilon^{AO}_{l}$ on the location - is due to anisoplanatism. When using an AO with a single mirror and a WFS, only the seeing that is common to the whole FOV can be corrected over the whole FOV. The more distant the atmospheric turbulence is, the smaller the area of the FOV that shares the same wave-front aberrations (isoplanatic patch), and the smaller, therefore, the area that can be fully corrected. However, any corrections that are specific to one isoplanatic patch are unavoidably applied to the whole FOV, and therefore added to the aberrations everywhere else, thus causing a statistical increase in the wave-front errors there, rather than a decrease. To avoid this, a ground-layer WFS can be used that relies on a larger area than one isoplanatic patch to calculate the wave-front slope, so that the cross-correlations are less sensitive to wave-front errors that are not shared across that area.\par

    In practice, this method reduces, but does not completely eliminate, the sensitivity to anisoplanatic wave-front aberrations, so that the correction of the wave-front phase errors becomes increasingly ineffective with distance from the region to which the AO is locked onto (hotspot). This leads to statistically poorer corrections at the edges of the FOV. A combined effort of evaluating how the corrections get worse with distance, and how the high altitude seeing intrinsically varies within the FOV is required in order to characterise this effect. In the context of this section, the most important realization concerning $\epsilon^{AO}_{l}$ is that they are generally unknown. This leaves the profile of residual wave-front phase errors unknown, and therefore the corresponding residual transfer functions unknown. Revising the image formation model to allow AO-compensations, and denoting the residual transfer functions corresponding to $\phi^{AO}$ as $\hat{T}^{AO}$, we write:

    \begin{equation}
       \hat{I}(\vec{k}, t) = \hat{T}(\vec{k}, t)\hat{O}(\vec{k}) + \hat{N}(\vec{k}).    \label{7}
    \end{equation}

    \noindent It is implicit that $\hat{T}(\vec{k}, t)$ corresponds to the combined optical system where the time-dependence is solely caused by the atmosphere. The subscripts have, therefore, been dropped. Further improvements to the quality of observations are availed with the use of post-facto processing methods such as MFBD\citep{vanNoort05, vanNoort17} that do not require prior knowledge of the degrading transfer functions. In MFBD image restoration, estimates of the transfer functions are made directly from the degraded observations. This is achieved by expanding the wave-front phase errors onto an orthogonal basis and fitting for the amplitude of every mode. However, the basis used by MFBD is ultimately finite and the expansion is truncated to a mode limit $M$, which causes a systematic underestimation of the wings of the Point Spread Functions (PSFs). Therefore, while MFBD corrects for degradation caused by lower-order modes, the effects of the higher orders (stray light) are left largely untreated. Since image restoration with MFBD is normally performed on observations that are already AO-compensated, one may assume that:

    \begin{align}
    \epsilon^{AO + MFBD}_{l} = 1, \hspace{10pt}\forall \hspace{10pt}l < M,   \label{8}
    \end{align}
    
    \noindent that is, residual degradation in AO-compensated MFBD-restored observations are exclusively due to the higher-order modes. A peculiar state of compensation is realised when the AO corrects more modes than MFBD ($L > M$). In this case, modes with $l$ < M are perfectly compensated by the AO and MFBD, modes with M < $l$ < L have been compensated by the AO (but with unknown efficiency), and modes with $l$ > L have not been compensated at all. Alternatively, one may try to perform image restoration with $M$ increased to \textasciitilde $L$ so that the issue of the unknown correction efficiency of the AO is circumvented entirely. However, given that the efficiency of wave-front corrections made by the AO vary within the FOV, mitigation of this issue is attempted during image restoration - by segmenting the FOV into a grid of patches and restoring each patch individually. In doing so, the assumption that wave-front phase errors over each patch are isoplanatic is made. This imposes a limitation on the admissible physical size of the PSFs, and indicates that restoring with arbitrarily high mode limits is not beneficial. The preferable approach is therefore to develop a self-consistent method of removing residual stray light from the restored data directly. \par
%
    
    
    We clarify that the term 'residual stray light' here refers exclusively to the component of stray light arising from high-order aberrations. Other sources of stray light, such as true scattering in the atmosphere and within the instrument, are beyond the scope of this paper. Correspondingly, we refer to the PSF describing stray-light contamination as a stray-light PSF.
    
\section{Residual degradation in slit spectra}
    \label{Sec.3}

    In the previous section, we reviewed the limitations of AO and MFBD and qualitatively addressed the nature of residual degradation in ground-based observations. We now describe the characteristics of residual stray-light contamination in spectropolarimetric data obtained with Spectrographs. A detailed discussion on how such data are obtained and restored is found in \citet{vanNoort17}. Here, we only briefly review the relevant details. In the interest of keeping the description concise, we ignore the polarimetric aspect of such data. We assume that the conclusions drawn in this section apply to all the Stokes components equally. \par

    Slit-spectra are obtained with a detector that collects light that has passed through a slit, and that has been dispersed by a dispersive element, usually a grating. Another detector (slit-jaw camera) is used to image the slit and the surrounding FOV through a broadband filter spanning the spectral region covered by the spectra. Both the detectors are operated synchronously so that the recorded images/spectra can be assumed to be degraded by the same seeing, and the frame-rates of the detectors are kept high enough to prevent seeing-induced distortions, and seeing-induced crosstalk. The images recorded by the slit-jaw camera are restored in bursts with MFBD to yield estimates of the degrading PSFs as a function of space and time. However, to restore the slit-spectra, one cannot directly deconvolve them as the intensities perpendicular to the slit have not been recorded co-temporally. Instead, the problem is reformulated into a set of linear equations that connect the observed intensities with the undegraded solar spectra through the PSFs. That is, Eq.1 can be re-written into a matrix equation of the form: 

    \begin{equation}
        \hat{M} \textbf{x} = \textbf{y}, \label{9}
    \end{equation}

    \noindent where $\textbf{x}$ represents the coveted undegraded spectra, $\textbf{y}$ represents the seeing-degraded intensities recorded by the spectral camera, and $\hat{M}$ is, in principle, a non-sparse non-square matrix that represents an instantaneous PSF. The values of such PSFs are obtained from the MFBD restoration of the slit-jaw images, and therefore the matrix $\hat{M}$ is completely known. Eq.~\ref{9} can be constructed for every wavelength point, and for each instant of time, yielding an over-determined system of coupled equations. Since $\hat{M}$ is not a square matrix, one could instead solve the pseudo-inverse problem:
    
    \begin{align}
        \hat{A}\vec{x} = \vec{b}, \label{10}
    \end{align}

    \noindent where $\hat{A} = \hat{M^{T}}\hat{M}$ is a square matrix, and $\vec{b} = \hat{M^{T}}\vec{y}$. However, since a large number of equations and unknowns are typically involved, it is difficult to directly invert the matrix $\hat{A}$. The solution therefore, is obtained using a variation of the iterative Lucy-Richardson scheme. Since the matrix $\hat{A}$ is constructed using PSFs whose wings are underestimated, the solution thus obtained is likely to still contain some residual stray light. If we were to estimate a correction $\delta \hat{A}$ to $\hat{A}$ that addressed this issue, the most self-consistent way forwards would be to construct a new matrix $\hat{A'} = \hat{A} + \delta\hat{A}$ and restore the raw data again. However, due to the rapid scaling of computational workload with the size of the domain, using $\hat{A'}$ becomes prohibitively expensive. We therefore propose an alternate method of removing stray light directly from the restored spectra. 
    
    First, the Stokes signals are converted to modulated intensities using an arbitrary modulation scheme for each wavelength, yielding four modulated intensities corresponding to each Stokes parameter. This is followed by degrading the modulated intensities with a diffraction limited PSF corresponding to the aperture of the telescope. The degraded intensities corresponding to each Stokes parameter are then deconvolved using the same set of PSFs that describe residual stray light exclusively across all wavelengths. Although the deconvolution is bound to amplify the noise in the signals, this amplification is uniform across the Stokes parameters and does not, therefore, introduce spurious polarization signals(polarimetric precision is preserved). After the deconvolution is performed, the same modulation scheme used in the forward modulation is used to invert the deconvolved intensities back into Stokes spectra. The method described above requires that a set of suitable stray-light PSFs be found. Two key aspects of such a set of stray-light PSFs have to be noted.
    
    Firstly, at any given wavelength, slit-spectra are not truly images because the intensities along the scan-direction have been recorded at different times. It then follows that residual stray light generally varies along the scan-direction as the seeing evolves with time. Therefore, to remove residual stray light from a single 'image' at a particular wavelength, multiple stray-light PSFs spanning the duration of the scan have to be computed. This complication is not shared by observations made with filtergraphs, as intensities there are obtained strictly co-temporally. Secondly, consider a fixed point on the Sun. Let the slit sample this point at time $t_{0}$. According to Eq.~(\ref{10}), the true solar intensity of this point (solution) is computed from the intensities in the slit recorded between $t_{0} - \Delta t$ and $t_{0} + \Delta t$, where $2\Delta t$ is the time it takes for the slit to scan the spatial extent of the PSF. Naturally, intensities recorded at $t_{0}$ carry the highest weights towards the solution while those recorded at $t_{0} \pm \Delta t$ carry the lowest. The correct amount of residual stray light is therefore described not just by one stray-light PSF computed at $t_{0}$, but through a weighted average of many stray-light PSFs computed between $t_{0} - \Delta t$ and $t_{0} + \Delta t$, with the weights specified in the matrix $\hat{A}$. \par
    
\section{Estimating the stray-light PSF}
    \label{Sec.4}

    \begin{figure*}
    \includegraphics[width=\textwidth]{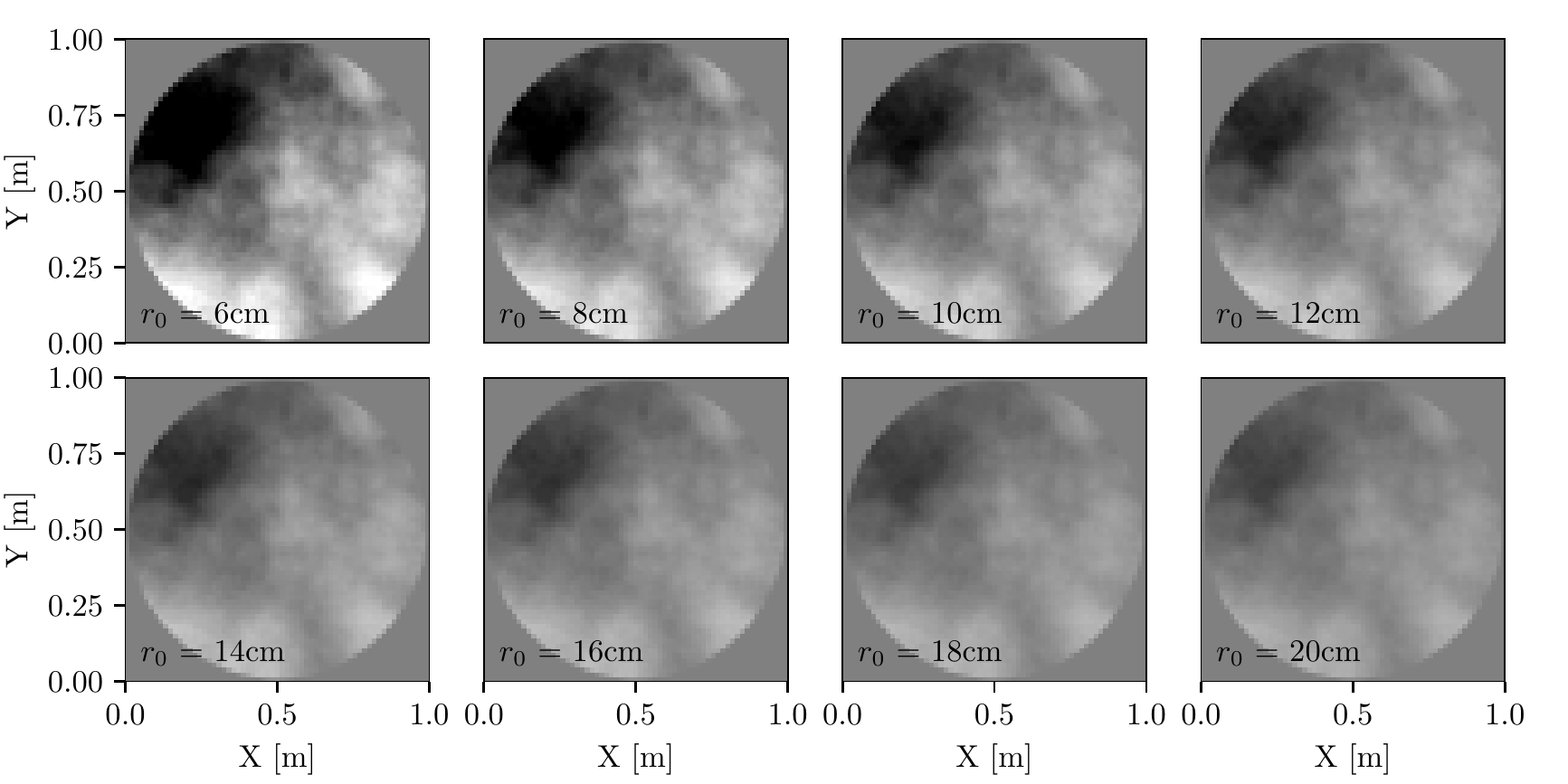}
    \caption{Kolmogorov wave fronts simulated for different values of the Fried parameter \Fried{}, for an aperture size of 1 m. The mean values of all the wave fronts have been subtracted, and a uniform grey scale has been used in the illustration of the wave fronts.}
    
    \label{Fig1}
    \end{figure*}

    \citet{Scharmer10} followed a statistical method, where values of the Fried parameter (\Fried{}) were used to simulate a set of wave fronts obeying Kolmogorov statistics. \Fried{} values were calculated using a setup employing a wide-field wave-front sensor installed before the tip-tilt mirrors and the DM. At the time, the DM was driven by a 37-electrode configuration, corresponding to roughly 37 KL-modes. However, the DM at SST has since been upgraded to a configuration using \textasciitilde 85 electrodes, or roughly 100 KL-modes. As mentioned earlier, image-restoration is typically performed using MFBD with a mode limit of \textasciitilde 40, so that the amplitude of KL-modes between $l = 40$ and $l = 100$ in the residual wave fronts is unknown. We propose that the average correction efficiency of the AO can be computed from the following procedure: (i) the \Fried{} values are first used to simulate wave fronts obeying Kolmogorov statistics; (ii) the corresponding transfer functions computed through Eq.~(\ref{2})~\&~Eq.~(\ref{3}) are used to degrade an image synthesised from MHD simulations; (iii) both the sets of observations(synthetic and actual) are restored using MFBD (which returns the mode amplitudes); (iv) the ratio of the standard deviations of the mode amplitudes (Eq.~\ref{11}) in both the restorations yields the efficiency of the AO as a function of the mode. This efficiency is mathematically expressed as:

    \begin{align}
            \epsilon^{AO}_{l} = 1 - (\sigma_{l}^{obs}/\sigma_{l}^{syn}),    \label{11}
    \end{align}
    
    \noindent where $l$ denotes the mode number, and $\sigma^{syn}_{l}$ and  $\sigma^{obs}_{l}$ are standard deviations of the mode amplitudes returned by MFBD restoration of the synthetic and original recorded observations, respectively. Interpretation of Eq.~(\ref{11}) is straightforward for the trivial cases. For example, if the recorded observations are found to not have any degradation caused by mode $l$, the standard deviation for mode $l$ computed from image restoration would ideally be zero. This then implies that the AO perfectly compensated mode $l$ in the incident wave fronts ($\epsilon^{AO}_{l} = 1$). \par
    
    A potential source of error in the calculation of $\epsilon^{AO}_{l}$ is the assumption that MFBD works identically on the input observations, irrespective of the state of their degradation. To address this issue, the procedure outlined above can be followed iteratively - after the first round of $\epsilon^{AO}_{l}$ are found, these can be used to subtract modes from the incident wave fronts. These partially compensated wave fronts can then be used to generate another set of degraded observations followed by another round of image-restoration, and the cycle can be followed until the computed efficiencies are found to converge. Once a converged set of $\epsilon^{AO}_{l}$ are found, residual stray light in the observations is estimated by first perfectly subtracting all modes with $l < M$, followed by subtracting modes with $l > M$ with efficiency $\epsilon^{AO}_{l}$.
%
%
    The residual wave fronts thus generated are largely dominated by higher-order aberrations, and they approximate the amount of residual stray light still present in the observations. \par

\section{Kolmogorov phase screens}
    \label{Sec.5}

    A frequent approach to characterizing AO-systems is through the use of Kolmogorov phase screens, which are simulations of the spatial distribution of phase errors (wave fronts) incident on the telescope. These simulations are statistical in nature since modelling the exact propagation of a wave front requires knowledge of the refractive index at all spatial locations within the path of the wave front. The statistical properties of these Kolmogorov phase screens are well established in the works of \citet{Noll76}, \citet{McGlamery76}, \citet{Lane92}, and \citet{Glindemann93}. They conform to the Kolmogorov-Obukhov power law, and can be numerically simulated \citep{Nagaraju12} as a function of \Fried{}:

    \begin{align}
        \Gamma(\vec{k}) = 0.023 (D/r_{0})^{-5/3} |\vec{k}|^{-\beta}, \label{12}
    \end{align}

    \noindent where $\Gamma(\vec{k})$ is the power at spatial frequency $\vec{k}$, $D$ is the diameter of the phase screen, and $\beta = 11/3$. Fig.~\ref{Fig1} shows examples of such phase screens simulated over a circular aperture of $1$ m diameter. The mean value of all wave fronts has been subtracted, but no other modes have been removed. The phase screens are then used to compute the PSFs with which the synthetic MHD-generated images are degraded, and an appropriate amount of noise, by comparison with slit-jaw observations for example, is added to the synthetic data. We make a note that the validity of the MHD simulations is irrelevant as we are only interested in the synthetic images. Restoring the synthetic data and the original recorded observations with MFBD facilitates a comparison of their respective mode amplitudes.

\section{Observations and data reduction}
    \label{Sec.6}

    \begin{figure*}
    \includegraphics[width=\hsize]{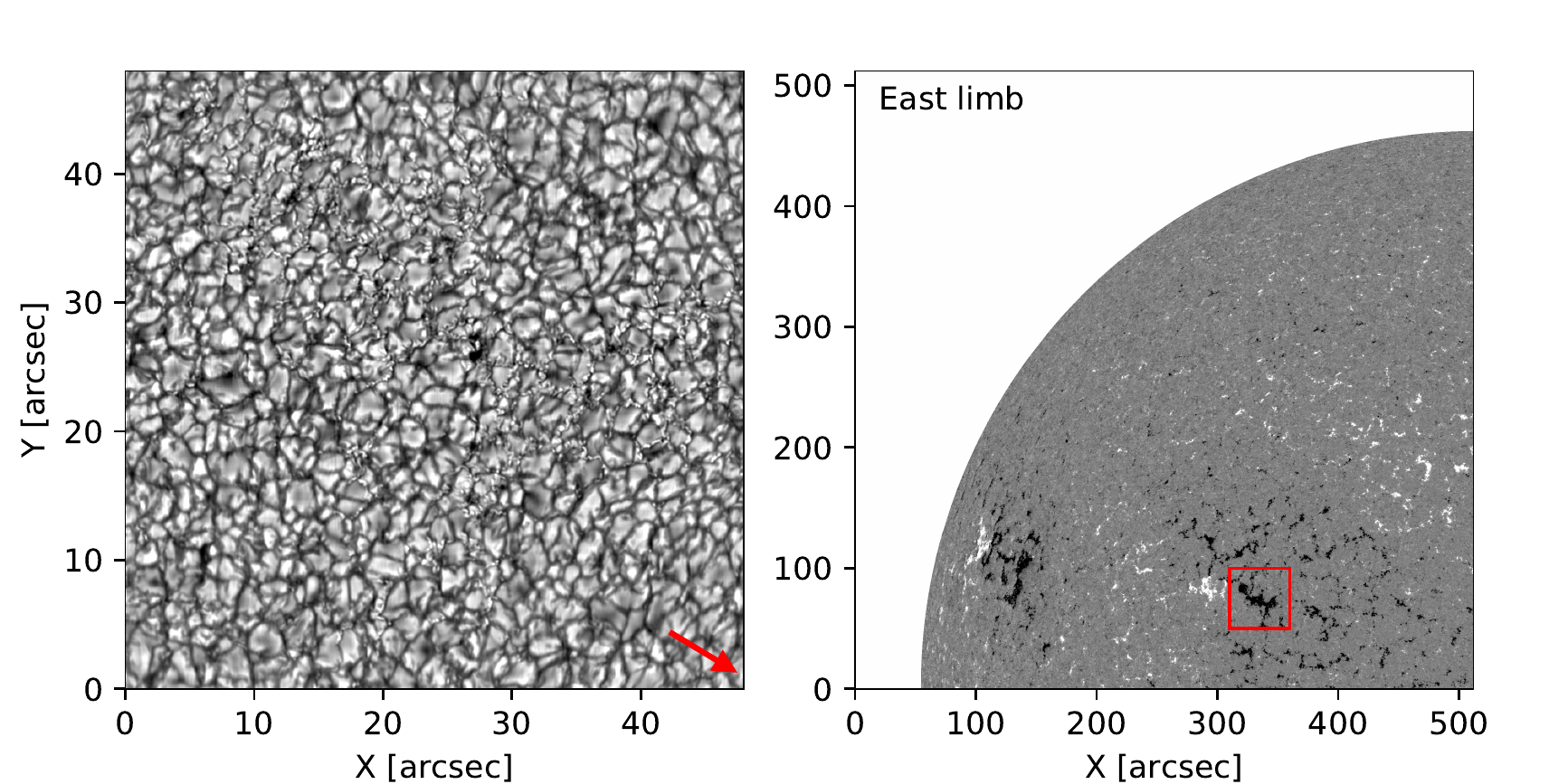}
    \caption{Panels showing the FOV and its location with respect to the full solar disk. \textit{Left}: Continuum intensity image from the restored observations. The red arrow points towards the disk centre. \textit{Right}: Line-of-sight magnetogram from HMI taken at approximately the same time. The red box highlights the FOV of the observations.  
    }
    \label{Fig2}
    \end{figure*}
       
    \begin{figure}
    \includegraphics[width=\columnwidth]{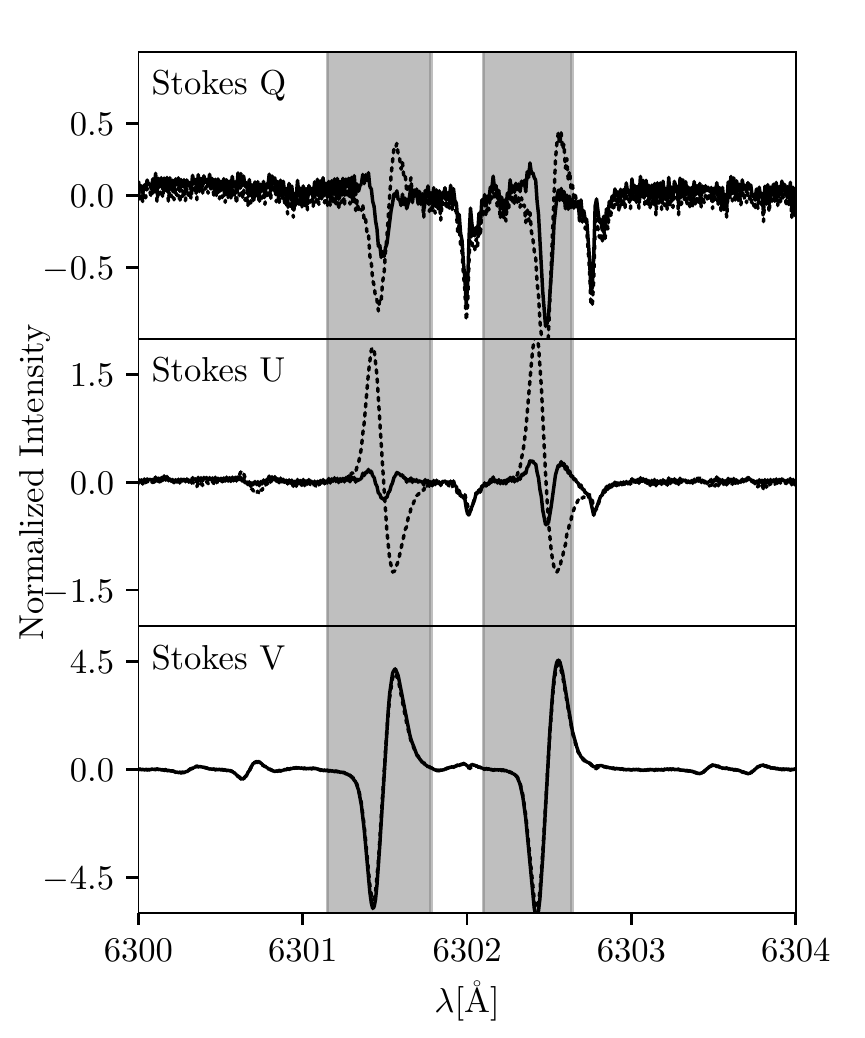}
    \caption{
        \textit{Top to bottom:} Stokes $Q$, $U$, and $V$ profiles averaged over the FOV. The dotted lines and solid lines show the spectra before and after crosstalk correction, respectively. The grey shaded areas indicate the wavelengths over which the dot products are computed. The units of intensity are arbitrary, and the profiles are not normalised with respect to Stokes $I$.}
    \label{Fig3}
    \end{figure}

    The dataset used in this paper was obtained on 28 June 2016 at the SST \citep{Scharmer03}, using the TRIPPEL spectrograph, a detailed description of which has been provided by \citet{Kiselman11}. A fast polarimeter designed for slit-spectra - based on the concept of FSP \citep{Iglesias16} - was installed and used during the same campaign (TRIPPEL-SP, Doerr et. al 2019, in prep). The scan was made under moderate seeing conditions with the help of an AO consisting of an 85-electrode bi-morph mirror, a Shack-Hartmann wave-front-Sensor (WFS), a correlation tracker, and a separate mirror for tip-tilt adjustments. The AO reported an average Fried parameter value of $13$ cm over the duration of the scan.

    Two Jai SP20000 CMOS cameras were used for recording the slit-jaw images, whereas the spectra were recorded using two Jai SP12000 CMOS cameras. The latter allowed 4-Mpixel frames to be recorded at a rate of up to $400$ Hz, the designed modulation rate of the FSP modulator. All cameras were synchronised using an external trigger signal. Although frames were recorded with a spatial sampling of $0.044$\arcsec, the sampling of the restored slit-spectra was scaled down to $0.06$\arcsec - corresponding to the critical sampling of the SST at $6300$ \AA{}. \par

    \subsection{Image restoration}

    The slit-jaw frames were restored in bursts of 2000 frames (5 s) with a temporal overlap of 3.75 s. As indicated in earlier sections, the restoration was performed over a grid of patches that were 128 $\times$ 128 pixel$^{2}$ in size, with an overlap of $118 \times 118$ pixel$^{2}$. The PSFs returned by MFBD for each patch were interpolated across the length of the slit. Karhunen-Lo\`eve modes were used as the basis functions, and the mode limit for the restorations was set to $M=45$. As outlined in Sec.~(\ref{Sec.3}), the values of the PSFs were used to construct the matrix $\hat{A}$, and subsequently restore the data recorded by the spectral camera.\par

    The restored slit-spectra cover a 53\arcsec $\times$ 53\arcsec FOV around a mostly uni-polar active region Plage, which is located towards the east limb at a heliocentric angle of 23 degrees ($\mu = 0.92$). The left panel in Fig.~\ref{Fig2} shows a Stokes I image from the restored dataset, extracted at a continuum wavelength near $6296.0$ \AA{}. The spectral window of the observations ranges from 6288.0 \AA{} to $6304.0$ \AA{}, with a sampling of 8.6 m\AA{}. The wavelengths were calibrated by fitting the flat-field (obtained at disk centre) using a least-squares method to the standard FTS spectrum from \citet{Delbouille73}. A total of 16 usable solar spectral lines were identified in the spectral window, including the iron pair at $6301.5$ \AA{} and $6302.5$ \AA{}. On the panel in the right of Fig.~\ref{Fig2}, a LOS magnetogram from HMI \citep{Scherrer12} is displayed for context.

    \subsection{Crosstalk and grey stray light}

    \begin{figure*}
    \includegraphics[width=\textwidth]{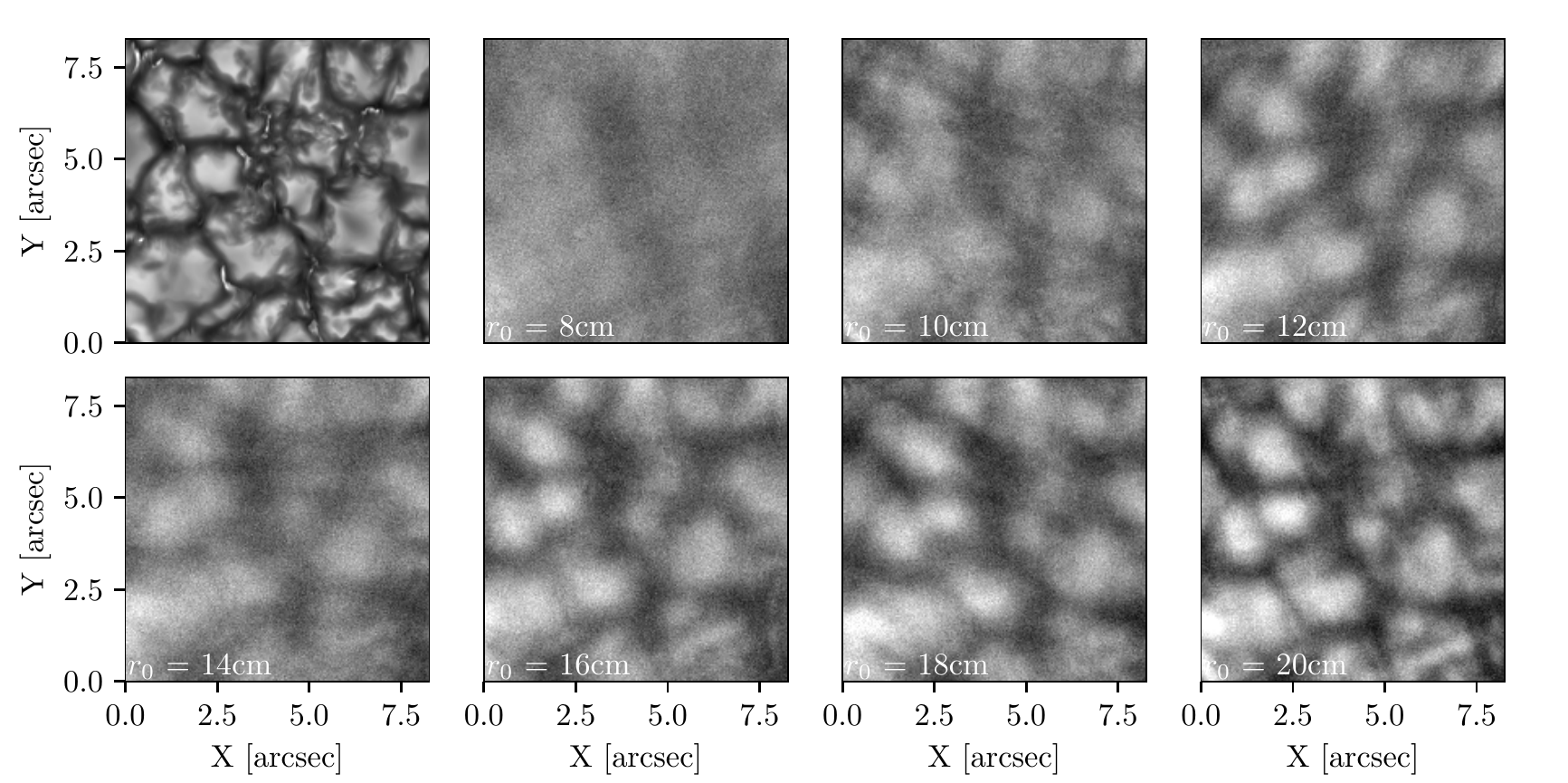}
    \caption{Panels showing degraded images synthesised from an MHD simulation at a continuum wavelength near $6300\AA{}$ for different values of \Fried{} (mentioned in the panels). The grey scales of the degraded images are identical. The grey scale range of the undegraded image has been enlarged to avoid extreme contrasts. The sampling in all images is $0.044\arcsec$. \textit{Top left}: Undegraded image (no added noise) corresponding to \Fried{} = $\infty$.}
    \label{Fig4}
    \end{figure*}
    
    The restored images indicated the presence of crosstalk from $I\xrightarrow{}Q, U, V$ and from $V\xrightarrow{}Q, U$, a consequence of neglecting the polarizing effects of the telescope optics so far. Although models of the Mueller matrix of SST exist, we found it convenient to apply ad hoc corrections to the spectra. The crosstalk from $I\xrightarrow{}Q, U, V$ was corrected by subtracting the average offsets (averaged over the FOV at continuum wavelengths) of Stokes $Q, U$, and $V$ (normalised with respect to Stokes $I$). To remove crosstalk from Stokes $V$, we parameterised the Stokes vectors on the Poincar{\'e} sphere, and rotated the Stokes components such that the FOV-averaged Stokes $Q, U$, and $V$ were symmetric, symmetric, and anti-symmetric in wavelength with respect to the line-centres, respectively. In doing so, we assume that the core of the plage makes the dominant contribution to the Stokes $V$ spectra, and that in these regions, gradients in the line-of-sight velocities and magnetic fields are not too strong to produce significant intrinsic asymmetry. The shaded region in Fig.~\ref{Fig3} indicate weighting of the profiles to include only solar lines in the (anti)symmetrization procedure. Since most of the crosstalk is from $V\xrightarrow{}Q, U$, the linear polarization signals look significantly different after crosstalk removal, while Stokes $V$ remains largely unchanged (Fig.~\ref{Fig3}).\par
    
    To remove the wavelength-independent component of the scattered light possibly caused by the extended wings of the spectral PSF (spectral veil, \citep{Borrero2016}), we make the assumption that the spectral PSF of the instrument is approximated to first order by a gaussian kernel with an initial guess value for its Full-Width Half Maxima (FWHM). The flat-field spectra obtained at disk centre is then expressed as the sum of the standard FTS spectra convolved with this gaussian kernel, and the additive spectral veil (expressed as a fraction of the continuum intensity). We then retrieved, both the FWHM of the gaussian kernel and the additive component, by iteratively fitting the flat-field spectra using a $\chi^{2}$ minimization method. The FWHM was found to be 41m\AA{} and the fraction of spectral veil was found to be 5\% of the continuum intensity.

    \subsection{Fried parameter}
    
    \begin{figure*}
    \includegraphics[width=\textwidth]{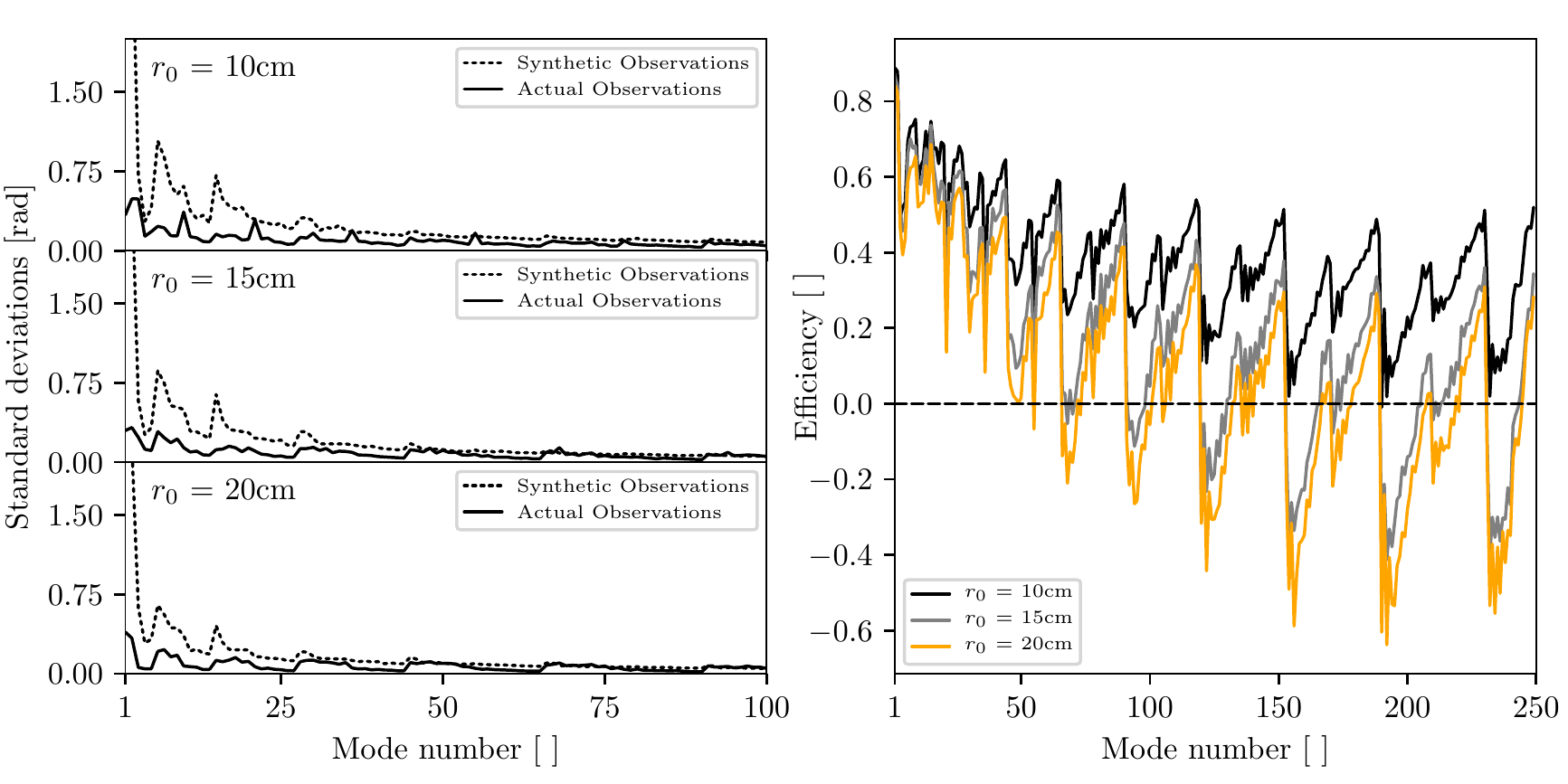}
    \caption{Plots of quantities obtained from MFBD restorations. \textit{Left:} Standard deviation of mode amplitudes plotted as a function of mode number. Mode amplitudes were obtained from MFBD restoration of degraded images. Only up to 100 modes are shown here. \textit{Right}: Average efficiency of the AO obtained from Eq.~(\ref{11}) plotted for different different values of \Fried{}}
    \label{Fig5}
    \end{figure*}

    \begin{figure*}
    \includegraphics[width=\textwidth]{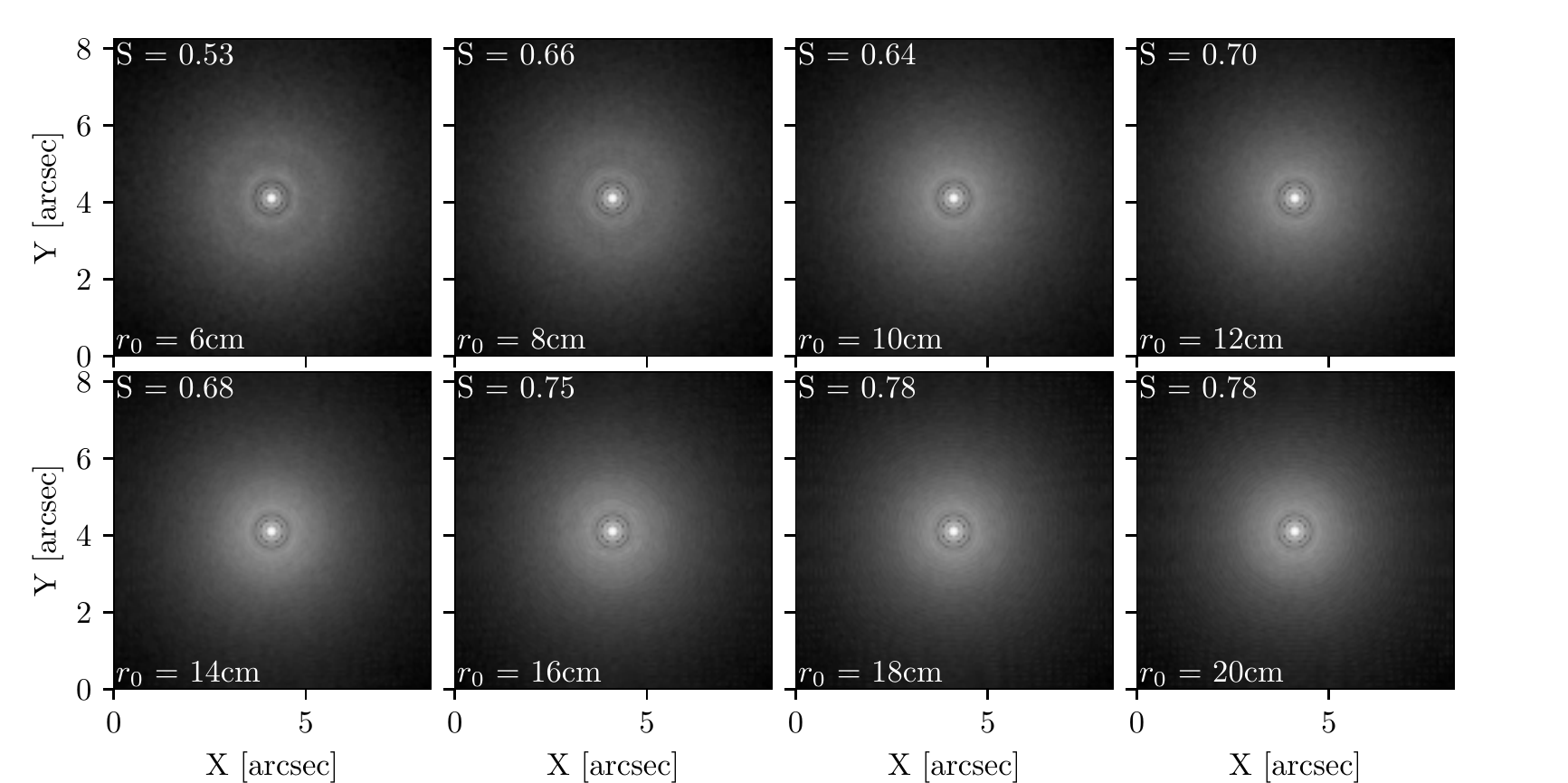}
    \caption{
        Stray-light PSFs for different \Fried{}, computed from the residual phase screens. The Strehl ratios of the PSFs are indicated in top left of each panel. At \Fried{} = 20 cm the radius encircling 90\% energy is around 1.1\arcsec.}
    \label{Fig6}
    \end{figure*}
    
    In closed-loop operation of the AO at the SST, an \Fried{} value for every second is computed from the voltages output to the DM-electrodes, as well as the variance of residual shifts of the $24\times24$ pixel$^{2}$ sub-images in four select sub-apertures, similar to a Differential Image Motion Monitor \citep{Sarazin90}. Further, another concurrent set of \Fried{} values are also reported using the same method, but with smaller $8\times8$ pixel$^{2}$ sub-images. Since cross-correlations between larger images are more sensitive to shifts that are global, the values of \Fried{} computed from the $24\times24$ pixel$^{2}$ sub-images are biased more towards ground-layer seeing that is shared by the full FOV. To avoid this bias, we use the values of \Fried{} computed from the $8\times8$ pixel$^{2}$ sub-images instead, and consider these to be more representative of seeing originating in both the ground and upper layers.\par

    \subsection{Calculating the stray-light PSF}
    \label{Sec.6.4}

    An ensemble of $2000$ Kolmogorov phase screens were created for each value of \Fried{} between $r_{0} = 6$ cm and $r_{0} = 21$ cm in steps of $1$ cm. A snapshot from an MHD simulation of a quiet Sun, performed with MURaM \citep{Vogler05}, was then used to synthesise an image at a continuum wavelength near $6300$\AA{} with the SPINOR radiative transfer code \citep{Frutiger00}. The synthesised image was subsequently degraded with the transfer functions corresponding to each of the $2000$ Kolmogorov phase screens simulated for each \Fried{}, followed by the addition of Poisson noise to each of the degraded images. The amount of added noise was calibrated from the power-spectra of the slit-jaw images in the recorded observations. Fig.~\ref{Fig4} shows a panel of these degraded images for the binned values of \Fried{}.\par
    
    We then setup parallel image restorations of the synthetic degraded data generated above, and the recorded observations. The mode limit used in the restorations was set to $M = 250$. Since MFBD is applied on AO-compensated data in practice, we avoided restoring the synthetic data for $r_{0} < 10$ cm. Further, due to the increased mode limit, the restoration of the recorded observations was constrained to the patch at the centre of the FOV. The implications of this restriction are that the stray-light PSFs that we compute would likely underestimate the residual stray-light contamination in the edges of the FOV. \par
    
    MFBD returns the fitted mode amplitudes for each KL-mode, and for each degraded image. This allowed us to compute their standard deviations, which are plotted in the left panel of Fig.~\ref{Fig5} for $r_{0} = 5, 10, 20$ cm. 
    The dotted and solid standard-deviation curves correspond to the synthetic degraded data, and the original recorded observations, respectively. Only the first $M = 100$ modes are shown to magnify the behaviour of these curves at the lower orders, but they continue to decrease all the way upto mode $M=250$. It is immediately seen that the standard deviations of the mode amplitudes in the recorded observations are lower than those corresponding to synthetic data, as it should be since the recorded observations are AO-compensated while the synthetic data are not AO-compensated. Another point to note from these curves is that the standard deviations of the synthetic degraded data decrease more rapidly than the recorded observations. This suggests that the AO-system at the SST works to keep the level of residual degradation in the observations fairly stable, irrespective of the seeing conditions. \par
    
    From the standard-deviation curves, we computed the efficiency of the AO-system as described by Eq.~(\ref{11}), for every second of the scan. The panel on the right of Fig.~\ref{Fig5} shows the computed efficiency of the AO ($\epsilon^{AO}_{l}$) as a function of the basis mode, averaged over the realizations for $r_{0} = 5, 10, 15$ cm. We now discuss the trends and features of these efficiency curves.\par
    
    The first is the presence of a periodic saw-tooth pattern, observable also in the standard-deviation curves although not as strongly. These peaks correspond to the radial KL-modes, and may be caused by the ordering of the KL-basis modes. Since the radial KL-modes correspond to fluctuations at a lower spatial frequency compared to the preceding and succeeding azimuthal modes, they would always have more power according to the Kolmogorov-Obukhov power law. However, these peaks should partly cancel each other out in the efficiency curves as efficiency is obtained as a ratio of standard deviations. As to why they are very prominent is not fully clear. A possible cause may be that the MFBD restoration is somewhat sensitive to the level of residual degradation in the data. An approach to circumvent this issue would be to use an iterative scheme, where AO-efficiencies at a given iteration are used to simulate AO-compensation on the fully uncompensated phase screens, and thereby generate another dataset. Upon restoring this dataset with MFBD again, and comparing the resulting standard deviations with the observations, one can arrive at a correction to the AO-efficiencies. The procedure can be repeated iteratively until the AO-efficiencies are found to converge. In our application, we noted that AO-efficiencies were usually overestimated at the very first iteration, and therefore the stray-light PSFs corresponding to this iteration likely underestimated the amount of residual degradation. However, upon subsequently deconvolving the data, we noted the amplification of artefacts in the spectra. The more aggressive the stray-light PSF, the larger the artefacts were amplified beyond the noise level. The properties of these artefacts are further discussed in Sec.~(\ref{Sec.7}). In order to avoid these artefacts, we therefore considered the stray-light PSFs obtained at the very first iteration to be final. We note that owing to this choice, we may slightly underestimate the amount of residual stray light in our slit spectra.\par
    
    The second is the indication that the AO continues to compensate very high-order modes consistently, particularly when the seeing is poor. While it is possible that certain higher-order modes are able to be compensated by the AO, the indication that modes upto $L = 250$ are corrected with an efficiency averaging $30 \%$ when $r_{0} = 5$ cm could also be due to theoretical errors in the variances of mode amplitudes in the simulated phase screens. For example, modifications to the exponent $\beta$ in Eq.~(\ref{12}) have been suggested in literature \citep{Dainty95, Rao00} with $2 < \beta < 4$, implying that the stray-light correction that we perform could be both an underestimation ($\beta = 2$) or an overestimation ($\beta = 4$). Another factor that affects the standard-deviation curves are errors in \Fried{}. Ideally, for a completely unbiased estimate of \Fried{}, the size of the sub-images in the WFS should be smaller, still, than $8\times8$ pixel$^{2}$. However, a lower bound exists as the AO cannot lock onto regions without sufficient intensity contrast. Consequently, the \Fried{} that is computed from the cross-correlation of these sub-images may deviate from the true value at any given time. The third feature of the AO-efficiency curves is that they also appear to take on negative values. Going back to the standard-deviation curves, this implies that the recorded observations have more degradation in the higher-orders than that caused by completely uncompensated Kolmogorov wave fronts! One possible explanation for this effect is that the AO-mirror adds spurious high-order aberrations to the incident wave fronts. This may be because the mirror cannot exactly reproduce the Zernike or KL-basis modes, leading to the amplification of high-order modes in the residual wave fronts. \par
    
    Despite some of the surprising results, the method outlines a self-consistent way of computing the performance of the AO-system from the recorded observations themselves. Since the slit spectra are AO-compensated and have been restored with MFBD, the amount of residual degradation in the slit spectra were estimated in the following way - first, modes until $M = 45$ were subtracted completely from the fully uncompensated phase screens. This step simulates the joint correction of AO and MFBD to the first $45$ KL-modes, which is assumed to be perfect. In the next step, modes from $M = 45$ to $M = 250$ were removed using the AO-efficiencies:
        
    \begin{align}
        \phi_{R} = \phi_{S} - \sum_{l=1}^{45}{C_{l}~Z_{l}} - \sum_{l=46}^{250}{\epsilon^{AO}_{l}~C_{l}~Z_{l}}.
    \end{align}
    
    \noindent Here, $\phi_{R}$ is the residual wave front, $\phi_{S}$ is the simulated wave front, and $C_{i}$ is the amplitude of mode $l$. The PSFs corresponding to the residual wave fronts were averaged over the realizations to yield a smooth stray-light PSF for every second of the scan. Following this, the stray-light PSFs were also running-averaged in time according to the weights specified in the matrix $\hat{A}$, and reduced to one stray-light PSF per row of the FOV (row is the scan-direction). These stray-light PSFs are displayed in Fig.~\ref{Fig6} for different values of \Fried{} in log scale. The Strehl ratios of the PSFs are indicated in the top left of each panel, and vary from 50\% in poor seeing to approximately 80\% in good seeing.

\section{Results and analysis}
     \label{Sec.7}

    \begin{figure*}
    \includegraphics[width=\textwidth]{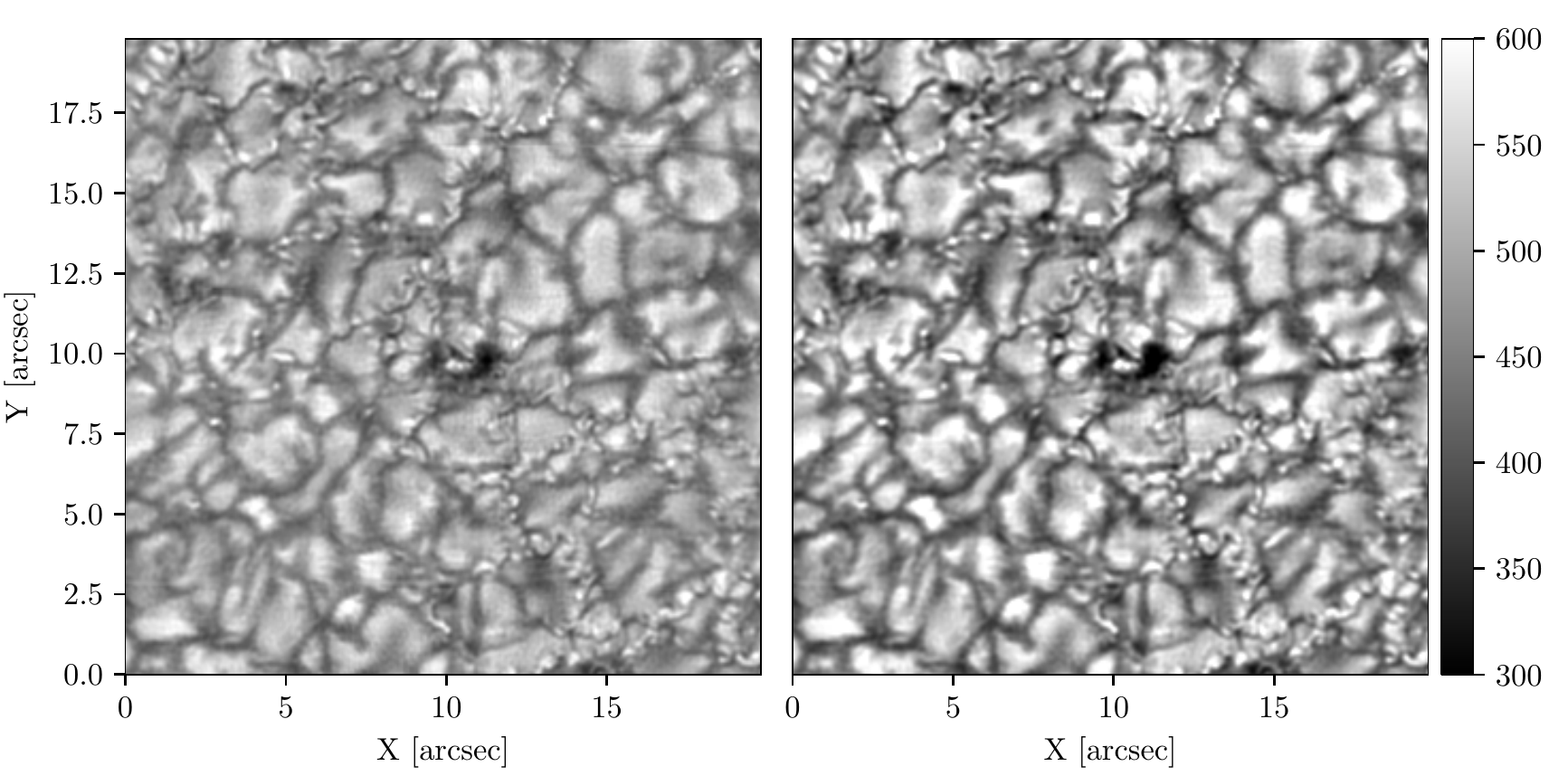}
    \caption{
      Comparison of the slit spectra at a continuum wavelength near 6300 \AA{}, before and after stray-light deconvolution. The grey scale has been kept uniform across both images, and the units of intensity shown here are arbitrary. The images shown here are at the centre of the larger FOV.}
    \label{Fig7}
    \end{figure*}
    
    Following the procedure outlined in Sec.~(\ref{Sec.3}), the Stokes profiles were first converted to modulated intensities using a balanced modulation scheme. The four modulated intensities were then subsequently deconvolved with an airy-disk PSF corresponding to the diffraction limit of the aperture of the SST, at each wavelength. To perform the deconvolution, the Lucy-Richardson scheme was used. The deconvolution was non-regularised and terminated after $20$ iterations as tests with synthetic data indicated that convergence is achieved quickly. We now discuss the effects of the stray-light removal procedure.\par
    
    \subsection{Effects on RMS granulation contrast}
    
    After deconvolution with the stray-light PSF, RMS granulation contrasts were found to increase to approximately to $12.3\%$ in the quiet Sun, from $9.0\%$ for the MFBD restored data. In Fig.~\ref{Fig7}, two continuum intensity images of the centre of the FOV are shown before (left) and after (right) the deconvolution. The grey scale has been kept equal so that the increase in contrast is readily visible in the deconvolved spectra. Since the method we followed works best at the centre of the FOV, the increase in contrast at the edges of the FOV is lower than it theoretically should be. \par
    
    In order to understand the reduction of stray light in our data, a comparison with synthetic images is required. We therefore synthesised an image at a continuum wavelength near $6300$ \AA{} (same region as our spectra) from an MHD simulation, and degraded it to $86.5\%$ of the diffraction limit of the telescope. This percentage was extracted from the power spectrum of deconvolved observations, and locating the spatial frequency above which noise dominates. Comparing the RMS contrasts, we found that a deficiency of $2.5-3\%$ still exists in our data. Since we are exclusively removing stray light due to higher-order aberrations, contributions from instrumental sources, and from atmospheric scattering still exist. However, as to how much they contribute to an increase in RMS contrast is still unclear.\par
    
    \subsection{Effects on spectra}
     
    \begin{figure}
    \includegraphics[width=\columnwidth]{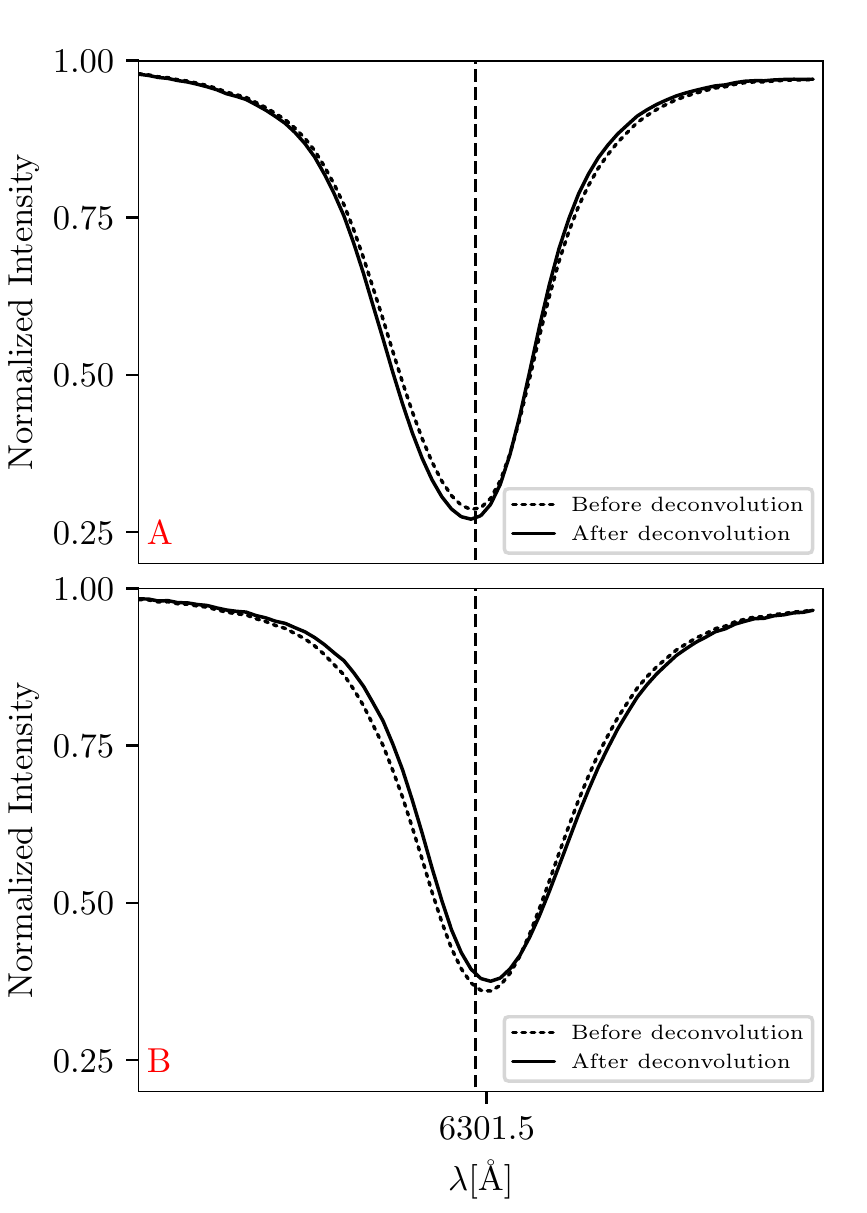}
    \caption{\textbf{A:} Comparison of the average line profile in granules before and after the deconvolution. Only the Fe I spectral line at 6301.5 \AA{} is shown here to magnify the differences. \textbf{B:} Comparison reproduced for the inter-granular lanes. The vertical scale for both plots have been kept the same.}
    \label{Fig8}
    \end{figure}
    
    \begin{figure}
    \includegraphics[width=\columnwidth]{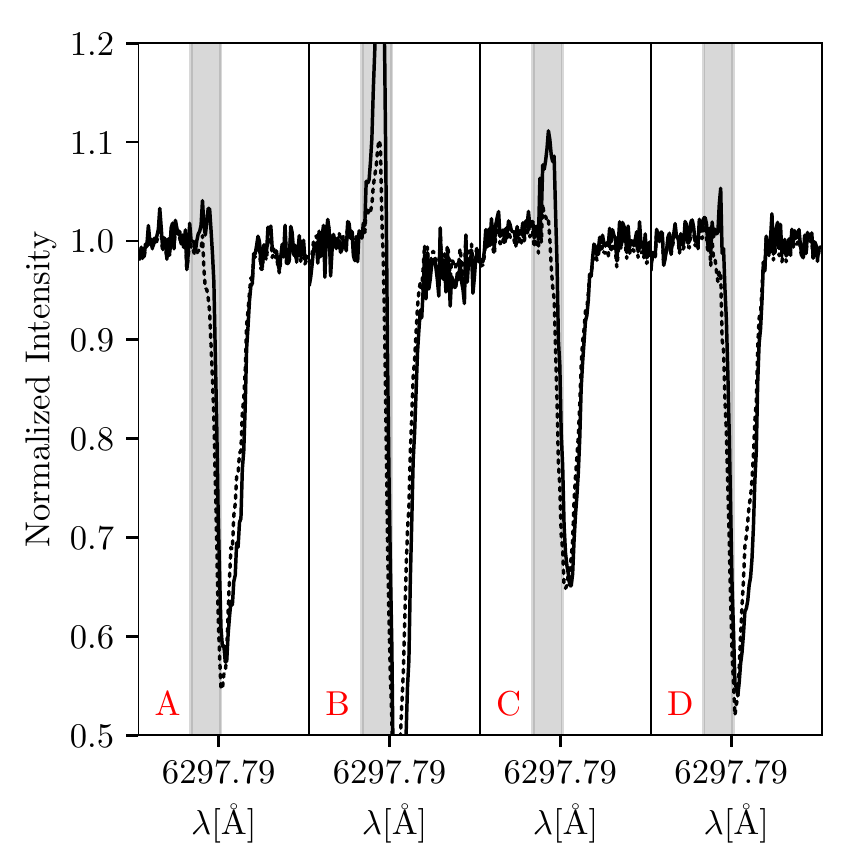}
    \caption{Line profiles of Fe I at $6297$ \AA{} shown before (dotted) and after (solid) the deconvolution in certain pixels. The grey shaded region highlights the location of the spike in the blue wing of the spectral line.}
    \label{Fig9}

    \end{figure}
    
    The deconvolution was found to affect the shape of spectral lines significantly. Fig.~\ref{Fig8} shows a comparison of the spatially averaged, normalised line profiles before and after the deconvolution in granules (top) and inter-granular lanes (bottom) for the Fe I line at $6301.5$ \AA{}. The selection of granules and inter-granular lanes was made on the basis of an intensity threshold. Although the differences are diminished due to the averaging, certain characteristic changes are visible. \par
    
    The depth of spectral lines in the granules was found to increase, as well as become more blue-shifted. This is because the stray-light profile in granules is composed of contributions from the surrounding inter-granular lanes which are comparatively shallower, and red-shifted due to the presence of down-flows. Subtracting these contributions, therefore, causes the spectral lines to become deeper and more blue-shifted in the granules. The converse holds true for the inter-granular lanes, where removing the stray-light contribution from granules causes the spectral lines within inter-granular lanes to become shallower and more red-shifted. These effects collectively indicate that temperature gradients, as well as line-of-sight velocities in both granules, and inter-granular lanes are intrinsically larger than what would be observed without stray-light correction.\par
    
    In the plage regions, the deconvolution was found to increase the amplitude of the Stokes $V$ lobes, while decreasing the depth of spectral lines in Stokes $I$. Since, Stokes $V$ is roughly proportional to the derivative of Stokes $I$ (under the weak-field approximation), deconvolution increases the deduced field strengths of the plage. However, several degeneracies in the atmospheric parameters affecting the shape of the spectral lines exist, and an accurate quantification of the effects of stray-light removal on spectropolarimetric data requires spectropolarimetric inversions.\par 
    
    The deconvolution was also found to amplify artefacts in certain pixels. Fig.~\ref{Fig9} shows a spectral line at $6297.79$ \AA{} in four different pixels (labelled A-D). The dotted and solid lines show the spectral line before and after stray-light deconvolution, respectively. In other words, the dotted line represents slit spectra that were restored with MFBD, and the solid lines represent slit-spectra that were restored with MFBD, followed by the stray-light deconvolution. The grey shaded regions indicate the location in the blue wing of the spectra line where a conspicuous spike appears. These spikes were found to be visible in all but the telluric lines, and their severity was found to vary. In some pixels (labelled B and C), the spike was already found in MFBD-restored data before the stray-light deconvolution. A likely cause for this may be the differential aberrations between the slit-jaw and the spectral cameras - when the slit moves or vibrates within the slit-jaw cameras' FOV during the scan. This implies that the PSFs returned by MFBD, based on the slit-jaw images alone, may not be able to precisely describe the degradation in the spectra recorded by the spectral camera. In pixels labelled A and D, it is harder to ascertain whether the spike existed before the stray-light deconvolution. Nonetheless, the stray-light deconvolution appears to have amplified it. We found the number of pixels displaying such artefacts above the noise level to be less than $0.1\%$ of the FOV.
   
    \section{Discussions and conclusions}
    
    In this paper, we extended the method initially developed by \citet{Scharmer10} - using Kolmogorov statistics - to self consistently estimate the amount of residual stray light directly from the observations. Although we focused on slit-spectra, our method is also applicable to observations obtained with filtergraphs. We found that the RMS contrasts of granulation in our observations increased by roughly three percentage points - from $9\%$ to $12.5\%$. This is much closer to the values of contrasts found in MHD simulations. Since the RMS contrasts in MHD simulations themselves show an intrinsic variability of one percent, the discrepancy between our observations and simulations may be of the order of one to three percent.\par
    
    In our method, we made several deliberate choices that theoretically lead to a slight underestimation of the residual stray light in our data for reasons of expediency, but it is possible to relax some of these constraints. For example, the estimation of the stray-light PSFs can also be performed as a function of position along the slit. One would therefore have a set of fully space variant stray-light PSFs. This requires more computational resources as multiple independent MFBD restorations need to be performed by segmenting the FOV into patches, but it would conceptually be similar to the method we have implemented here.\par
    
    We note that our method only removes the component of stray light due to high-order aberrations. Stray-light contamination from instrumental sources, and optical scattering remain unknown in our observations. Progress, however, is being made in accounting for these components. \citet{Lofdahl12} used a setup with an artificial target - situated at the primary of the focus of the telescope - to ascertain the amount of stray light originating in the post-focus optics at SST. They reported that contributions to stray light due to ghost images and scattering may be up to a few percent, in addition to the dominant contribution coming from high-order wave-front aberrations originating in the AO bimorph mirror.\par
    
    Due to the lack of a comprehensive solution that accounts for all the components of stray light, our observations are still likely to contain some residual degradation. However, given that our method accounts for, and restores a good chunk of the missing contrast, we consider our observations to be suitable for spectropolarimetric inversions.\par
    
    \begin{acknowledgements}
    The authors would like to thank Peter S{\"u}tterlin, and Hans-Peter D{\"o}rr for their technical support and useful discussions. This project has received funding from the European Research Council (ERC) under the European Union’s Horizon 2020 research and innovation program (grant agreement No. 695075) and has been supported by the BK21 plus program through the National Research Foundation (NRF) funded by the Ministry of Education of Korea. The participation of S. Saranathan was funded by the International Max Planck Research School for Solar System Science.
    \end{acknowledgements}
   
    \bibliographystyle{aa}
    \bibliography{literature}
\end{document}